\documentstyle[12pt]{article}

\begin{document}

\begin{flushright}
IJS-TP-99/04\\
February  1999\\
\end{flushright}

\vspace{.5cm}

\begin{center}

{\Large \bf SUSY and Symmetry Nonrestoration at High Temperature}

\vspace{1.5cm}

{\large \bf Borut Bajc}

\vspace{.5cm}

{\it J.Stefan Institute, 1001 Ljubljana, Slovenia\\}

\vspace{1cm}

\end{center}

\centerline{\large \bf ABSTRACT}

\vspace{0.5cm}

The status of internal symmetry breaking at high temperature 
in supersymmetric models is reviewed. This phenomenon 
could solve some well known cosmological problems, such as the 
domain wall, monopole and false vacuum problems.

\section*{Introduction}

The fascinating idea of symmetry nonrestoration at high temperature 
has been proposed long ago \cite{w74,ms79}, but is still a subject 
of ongoing research (see for example \cite{s98,b98} for recent 
reviews). As we will see, apart from being interesting by itself 
the phenomenon of symmetry nonrestoration could also naturally solve 
some cosmological problems. At first sight such a possibility seems 
forbidden in supersymmetric models  due to the existence of a 
no-go theorem \cite{h82,m84}. I will show how this difficulty can be 
circumvented.

Let me first explain in detail the choice of the title, motivating in 
such a way the work in this field.

{\it Why symmetry nonrestoration at high temperature?} 
We know that the low-energy symmetry of the Standard Model is 
described by the group $H=SU(3)_c\times SU(2)_L\times U(1)_Y$. 
It is natural to expect that at some high scale $M_X$ ($\approx 
10^{15}-10^{16}$ GeV) the three gauge 
couplings unify into one coupling, i.e. that $H$ is only a 
subgroup of a larger grandunified simple group $G$ (for example 
$SU(5)$, $SU(6)$, $SO(10)$, etc.). One would naively expect that 
at $T\gg M_X$ the ground state of the universe was symmetric under 
this group $G$, while it is clearly asymmetric at today temperatures 
$T\ll M_X$. Technically this can be studied by an order parameter, 
which is for example in the case of $G=SU(5)$ the vacuum expectation 
value (vev) of the $24$-dimensional adjoint $\Sigma$. Thus one 
usually thinks that $<\Sigma>=0$ at high $T$ (symmetry $G$ restored), 
while $<\Sigma>\ne 0$ signals the symmetry breaking $G\to H$ at small 
$T$. This means that in between these two extremes a phase 
transition took place. Being $G$ a simple group and 
having the low-energy group $H$ a $U(1)$ (hypercharge) factor, 
GUT monopoles were created during this phase transition via 
the well-known Kibble mechanism \cite{k76}. However this would 
lead to a cosmological disaster \cite{p79}: even if the 
monopoles were created during the phase transition at a rate 
of one per horizon, 
the fact that they could not decay sufficiently fast (so their 
energy density scaling only as $T^3$ instead as $T^4$ as for 
light particles) would be enough to have approximately $16$ 
orders of magnitude more energy today in the form of monopoles 
than in the form of baryons (obviously such a universe would have 
closed itself long ago). A similar derivation for models 
with domain walls gives very similar results and an analogue 
problem \cite{zko74}.

Such cosmological problems can be trivially solved if one finds 
that there were no phase transtions, i.e. that the grandunified 
group $G$ (for example $SU(5)$) was spontaneously broken already 
at high temperature $T>M_X$ to its low-energy subgroup $H$ (the SM 
group $SU(3)_c\times SU(2)_L\times U(1)_Y$) \cite{dms95,ds95}. 
In spite of a claim to the opposite \cite{bitu98}, 
in the case of global symmetries (domain wall problem) the 
majority of nonperturbative results \cite{roos96,jl98} confirm 
the possibility of symmetry nonrestoration. 
The question is still open in the case of gauge symmetries 
(monopole problem), where next to leading order corrections 
are big and tend to spoil the possibility of a perturbative 
expansion \cite{bl95,gprv98}. This gives a reason more to look for 
new noncanonical ways of gauge symmetry nonrestoration at 
high temperature. As I will show later, such solutions can be 
found and, contrary to naive expectations, are even more 
natural in supersymmetric models.

Another possible solution to the monopole problem 
is that the low-energy subgroup $H$ 
does not contain a $U(1)$ factor below $M_X$ \cite{lp80}. In 
such a scenario electromagnetism would be spontaneously broken 
during a period of the history of the universe and would eventually 
get restored later. Anyway, both in the case that $G$ is not restored 
and in the case that $H$ does not contain a $U(1)$ factor, we have 
symmetry breaking at high $T$, a phenomenon which is counterintuitive 
but extremely interesting.

Of course, there are also other possible solutions of the 
monopole and domain wall problems. The most fancy is probably 
inflation. Obviously we still want to have inflation (for example 
we need the Higgs field to be homogeneous), but some constraints 
on possible inflationary models needed to solve the above 
cosmological problems can be relaxed. Two other interesting 
scenario were proposed recently: in \cite{dlv97} it was shown 
that unstable domain walls could sweep away monopoles, while 
in \cite{bm99} moduli fields could have diluted the original 
monopole density. 

It is important here to stress once more that we still want 
inflation to take place. After all it has still to solve the 
usual horizon problem, homogeneity, etc. The only suggestion here 
is that inflation does not solve the monopole problem. Only in this way 
there is a hope that some monopole could one day be detected 
and thus give a clear experimental signature of a GUT. In fact 
if inflation solves the monopole problem, it pushes the monopole 
number essentially to zero, while thermal production even in the 
case of symmetry nonrestoration (and thus no phase transition) can 
give a sensible but nondangerous monopole density today 
\cite{t82,dms95}. More important, inflation needs more than 
just GUT, while symmetry nonrestoration could be obtained  
at least in principle from minimal models. We will see later 
that in SUSY such a solution does not only exist, but it is 
also very natural.

{\it Why SUSY?} The historical reason to consider seriously 
supersymmetric models is the solution to the hierarchy problem; 
on top of that the minimal $SU(5)$ SUSY GUT predicts correct gauge 
coupling unification, which cannot be achieved in the 
nonsupersymmetric version of the Standard Model \cite{ej82}. 
There are two special reasons which make the study 
of gauge symmetry nonrestoration in SUSY models at high 
temperature especially interesting. 

The first one is the appearence of a new cosmological problem 
\cite{w82}, usually ignored, but very important. It comes from 
the fact that SUSY models have (at $T=0$) often more than one 
vacuum, which are degenerate and disconnected. For example the 
minimal $SU(5)$ with one adjoint has three degenerate vacua: the one with 
$SU(5)$ symmetry (call it vacuum 1), one with $SU(5)$ broken to 
$SU(3)\times SU(2)\times U(1)$ (vacuum 2) and a third one with 
the symmetry $SU(4)\times U(1)$ (vacuum 3). Obviously, at low energy 
(but above the scale $M_W$) our universe was in vacuum 2. 
If one assumes as usual that at high $T$ the 
GUT symmetry was restored, when going from $T\gg M_X$ to 
$T\ll M_X$ the ground state would remain trapped in the wrong 
vacuum with $SU(5)$ unbroken. The point is that the barrier 
between two vacua is huge (of order $M_X^4$), so that tunneling 
is essentially impossible. Again, the problem could be solved 
if also at $T\gg M_X$ the GUT symmetry was spontaneously 
broken to the SM gauge group.

The second special reason to make SUSY models interesting is that 
apparently the program of symmetry nonrestoration does not 
work. There exists in fact a no-go theorem \cite{h82,m84}: 
any internal symmetry in a renormalizable SUSY model gets 
restored at high enough temperatures. The generalization 
to nonrenormalizable (effective) SUSY models has not been 
proven in general, but there is good evidence that it applies 
also in this case \cite{bms96,bs96}, despite some previous claims on 
the opposite \cite{dt96}. But, as is usually the case with no-go 
theorems, it is possible to evade them finding some more 
general examples, which were not considered by the authors of 
\cite{h82,m84}. In the following I will describe two 
such possibilities.

\section*{Large external charge density}

The above no-go theorem considers only cases with vanishing charge 
density, so it can not be applied here. The essential idea was 
developed for nonsupersymmetric models long ago, but it was realized 
only in \cite{rs97} that it can be applied in SUSY models as well. 
Since here supersymmetry is not essential and does not change any 
conclusion, I will sketch only the nonsupersymmetric case. Consider 
a toy model with a $U(1)$ global symmetry with the potential

\begin{equation}
V_0=\lambda |\phi|^4\;.
\label{v0}
\end{equation}

\noindent
At high temperature, the leading contribution is given by the term 
\cite{kl72,w74,dj74}

\begin{equation}
\Delta V_T={T^2\over 24}\sum_i{\partial^2 V_0\over\partial\phi_i^2}=
{\lambda T^2\over 3}|\phi|^2\;.
\label{dvt}
\end{equation}

\noindent
At high charge density $n$ one has to introduce also a chemical 
potential $\mu$, which modifies the effective potential by 
further new terms \cite{hw82,bbd91}

\begin{equation}
\Delta V_\mu=-\mu^2|\phi|^2-{\mu^2T^2\over 6}+\mu n\;.
\label{dvmu}
\end{equation}

\noindent
The effective potential thus starts with 

\begin{equation}
V=\left({\lambda T^2\over 3}-\mu^2\right)|\phi|^2+...\;,
\end{equation}

\noindent
so that the vev of $\phi$ becomes nonzero as long as the 
chemical potential is large enough ($\mu_{crit}=T\sqrt{\lambda/3}$). 
Of course, the right way to proceed is to impose the condition 
$\partial V/\partial\mu=0$ and thus rewrite the potential 
in terms of the charge density $n$ instead of the chemical potential 
$\mu$. In such a way one can find the critical charge density

\begin{equation}
n_{crit}=\sqrt{\lambda\over 27}T^3\;.
\end{equation}

For $n> n_{crit}$ the vev of $\phi$ is nonzero and the 
global symmetry $U(1)$ is spontaneously broken, while for 
$n\le n_{crit}$ the same vev is vanishing, thus restoring 
the global symmetry in question. Since both $n$ and $n_{crit}$ scale 
as $T^3$ during the history of the universe, the condition for 
an internal symmetry to be broken remains invariant as long as 
the charge is really conserved.

The most attractive way of implementing the above scenario is 
to have a large lepton number in the universe \cite{l76,ls94,brs97}. 
Notice that a large lepton number in the universe is experimentally 
allowed. From primordial nucleosynthesis constraints and galaxy 
formation one gets an approximate limit $n_L\le 70 T^3$ for 
$T\gg M_W$ \cite{ks92}, which is an order of magnitude more than the 
critical density found in \cite{brs97}. Also, such a large lepton number is 
consistent with a small baryon number ($\approx 10^{-9}T^3$) 
since sphalerons are not operative when $SU(2)_L$ is spontaneously 
broken (similarly to the usual $T=0$ tunneling, it is exponentially 
suppressed). So the picture is consistent, but unfortunately 
one Higgs only in the SM is not enough to break completely the 
whole gauge group, leaving the system with electromagnetic 
gauge invariance at any temperature. Thus the SM itself does not 
possess the solution to the monopole problem. This difficulty 
can be however easily circumvented in the MSSM, where several 
boson fields can get nonzero vevs breaking partially or completely 
the gauge group and thus solve the monopole problem \cite{bs99}. 

In some type of GUT supersymmetric models the role of the 
lepton symmetry can be played by an R-symmetry \cite{brs98}. 
A large R-charge would be eventually washed out at a later 
stage of the evolution of the universe due to supersymmetry 
breaking, so that contrary to the case with the lepton charge 
no signal of any R-charge would remain today. Nevertheless 
such a scenario would postpone the creation of monopoles to 
a nondangerous era.

\section*{Flat directions at high temperature}

Recently a new idea for symmetry nonrestoration in 
SUSY models was proposed \cite{dk98,bs98}. At $T=0$ it is very 
common in supersymmetric theories to have flat directions. 
The point is that by definition they can be very softly coupled 
to the rest of the world (for example with higher dimensional 
nonrenormalizable terms), so that it is not difficult to find 
them out of thermal 
equilibrium at high temperature. This is exactly the case which 
was not considered by the no-go theorem: in \cite{h82,m84} all 
the fields were assumed to be in thermal equilibrium, i.e. to 
develop a positive mass term $+T^2|\phi|^2$ in the potential, 
see (\ref{dvt}). To get the idea, let me consider the toy 
model from \cite{bs98} with the superpotential and K\" ahler 
potential given by 

\begin{eqnarray}
W&=&{\lambda\over 3}q^3+{\phi^{n+3}\over (n+3)M^n}\;\;,\;\;n\ge 1\;,\\
\label{w}
K&=&q^\dagger q+\phi^\dagger \phi+a{(q^\dagger q)(\phi^\dagger \phi)
\over M^2}\;.
\label{k}
\end{eqnarray}

Obviously, for $M\to\infty$ and 
$\lambda\approx O(1)$, the field $q$ is in thermal 
equilibrium (with itself through its self coupling), while 
$\phi$ is a flat direction and not in thermal equilibrium with 
$q$ (actually it is a free field). For a finite $M$ (let me 
take $M=M_{Pl}$ for simplicity) $\phi$ is no more an exact 
flat direction, but let us assume that it is still out 
of thermal equilibrium, due to the smallness of its interaction. 
As we said before, $q$ gets a positive mass term at high $T$, 
which drives its vev to zero, $<q>=0$. Therefore in the evaluation of the 
effective potential the field $q$ is never present as an external 
leg, but it runs in the loops, so we will denote it as $\hat{q}$. 
The opposite happens to the out of equilibrium field $\phi$: it 
does not run in thermal loops, but can have a nonzero vev, so 
it can have external legs in a Feynman diagram. We will denote 
this background field as before with $\phi$. 

The bosonic part of the Lagrangian is

\begin{equation}
{\cal L}_B=\left(1+a{|\phi|^2\over M^2}\right)|\partial\hat{q}|^2-
{\lambda^2|\hat{q}|^4\over 1+a|\phi|^2/M^2}-{|\phi|^{2(n+2)}/
M^{2n}\over 1+a|\hat{q}|^2/M^2}\;.
\label{lb}
\end{equation}

The kinetic term in (\ref{lb}) is not in the canonic form, so 
we have to rescale $\hat{q}\to\hat{q}/(1+a|\phi|^2/M^2)^{1/2}$. After 
expanding the Lagrangian in inverse powers of $M$, one gets the 
most important interaction part

\begin{equation}
{\cal L}_{int}=3a\lambda^2{|\hat{q}|^4|\phi|^2\over M^2}+...\;,
\end{equation}

\noindent
which generates nothing else than the ``butterfly'' diagram of 
\cite{bms96}. However, its sign depends on the parameter $a$. 
After including also the fermion loops one gets 

\begin{equation}
V_{eff}=-{3a\lambda^2\over 32}{T^4\over M^2}|\phi|^2+
{|\phi|^{2(n+2)}\over M^{2n}}\;.
\label{veff}
\end{equation}

\noindent
Clearly, for $a>0$ the effective potential (\ref{veff}) has 
a minimum at 

\begin{equation}
<\phi>^{n+1}=\left({3a\lambda^2\over 
32(n+2)}\right)^{1/2}T^2M^{n-1}\;,
\end{equation}

\noindent
which gives $<\phi>\approx T$ for $n=1$ and even $<\phi>\gg T$ 
for $n>1$. 

It is important to stress again that 
such a $<\phi>\ne 0$ at high $T$ means 
symmetry breaking at high $T$, which could be a possible 
solution to the above mentioned cosmological problems 
of monopoles, domain walls and wrong vacuum. The essential 
point is that the temperature provides the necessary breaking of 
supersymmetry needed to give large vevs to the would have been 
flat directions.

Let me give few comments. First, the constraint $a>0$ 
can be generalized in 
different models to a constraint for the K\" ahler 
potential. It is enough that $\partial^2 K/\partial q\partial q^\dagger$ 
grows with $\phi$. Such a constraint is very natural and not at all 
difficult to achieve. Second, the last term in the superpotential 
(\ref{w}) is needed, since without it the vev of $\phi$ would 
tend to infinity. Finally, a more realistic case with $\phi$ a charged 
field under some gauge group was considered in \cite{dk98} with 
similar conclusions.

\section*{Summary}

I showed that generic cosmological problems are present in SUSY GUTs: 
the monopole problem and the wrong vacuum problem, while sometimes 
also the domain wall problem can be present. An appealing solution to 
all these problems is given by the symmetry nonrestoration at high 
temperature. I described two such possibilities: the presence of a large 
lepton (or some other charge) density or the use of out of thermal 
equilibrium flat directions. 

{\it Acknowledgments}
It is a pleasure to thank all the people that taught me 
the physics I have been describing in this short paper: 
Goran Senjanovi\' c, who introduced me in this beautiful and 
interesting subject, Toni Riotto, Alejandra Melfo and Gia Dvali. 
I thank the organizers for this inspiring conference in a nice 
environment and for financial support. This work and 
the long trip to California were supported by the Ministry of 
Science and Technology of Slovenia.

\end{document}